\documentclass[preprint]{aastex}
\usepackage{amsmath}
\usepackage{amssymb}
\usepackage{bm}

\shorttitle{Direction of the Local IMF}
\shortauthors{Swisdak, Opher, Drake, and Bibi}

\begin{document}

\title{The Vector Direction of the Interstellar Magnetic Field Outside the
Heliosphere}

\author{M. Swisdak}
\affil{IREAP, University of Maryland,
    College Park, MD 20740-3511}
\email{swisdak@umd.edu}

\author{M. Opher}
\affil{Department of Physics and Astronomy, George Mason
University, Fairfax, VA, 22030}

\author{J. F. Drake}
\affil{IREAP, University of Maryland,
    College Park, MD 20740-3511}

\and 

\author{F. Alouani Bibi}
\affil{Department of Physics and Astronomy, George Mason
University, Fairfax, VA, 22030}

\begin{abstract}

We propose that magnetic reconnection at the heliopause only occurs
where the interstellar magnetic field points nearly anti-parallel to
the heliospheric field.  By using large-scale magnetohydrodynamic
(MHD) simulations of the heliosphere to provide the initial conditions
for kinetic simulations of heliopause (HP) reconnection we show that
the energetic pickup ions downstream from the solar wind termination
shock induce large diamagnetic drifts in the reconnecting plasma and
stabilize non-anti-parallel reconnection.  With this constraint the
MHD simulations can show where HP reconnection most likely occurs.  We
also suggest that reconnection triggers the 2-3 kHz radio bursts that
emanate from near the HP.  Requiring the burst locations to coincide
with the loci of anti-parallel reconnection allows us to determine,
for the first time, the vector direction of the local interstellar
magnetic field.  We find it to be oriented towards the southern solar
magnetic pole.
\end{abstract}

\section{Introduction}

After crossing the solar wind termination shock in 2004 and 2007,
respectively, the Voyager 1 and 2 spacecraft discovered that the
downstream flows remain supersonic with respect to the thermal ions.
This configuration had been anticipated \citep{zank96a} and occurs
because most ($\approx 80\%$) of the solar wind energy upstream of the
shock transfers to a non-thermal population whose presence, although
not directly detected, can be inferred from extrapolation of the
available data \citep{decker08a,richardson08a}.  The shocked solar
wind can then simultaneously exhibit subsonic flow with respect to the
energetically dominant component of the plasma and supersonic flow
with respect to the thermal component.

Pickup ions likely form the bulk of this non-thermal population.
Interstellar neutral atoms, unaffected by electromagnetic fields and
possessing a velocity characteristic of the solar system's motion with
respect to the local interstellar medium (LISM), $\approx
25\,\text{km/s}$, can drift inside the termination shock where they
encounter the $\approx 400\,\text{km/s}$ solar wind.  When these atoms
ionize they suddenly come under the influence of the solar wind's
magnetic field, are ``picked up'', and join the wind's outward flow.
The high relative velocity of the neutral atoms becomes an effective
non-thermal temperature with $k_BT = mv^2/2 \approx 1\,\text{keV}$.
Although not directly observed at the termination shock, pickup ions
have been detected throughout the inner heliosphere.

Models suggest that the Voyager spacecraft will remain within the
heliosphere for ten to twenty years (see Table 1 of
\cite{opher06a}) before finally encountering the HP, the
boundary between the solar system and local interstellar space.  There
the interstellar magnetic field abuts the heliospheric field in much
the same way that, within the solar system, the fields of magnetized
bodies meet the interplanetary magnetic field at magnetopauses.
Magnetic reconnection often occurs at such interfaces and has been
observed at, among other locations, Earth, Mars \citep{eastwood08a},
and Saturn \citep{huddleston98a}.  By analogy, it is also expected to
occur at the HP \citep{fahr86a}.

However, the high energy content of the pickup ions implies a large
value for the heliosheath plasma $\beta$ (where $\beta = 8\pi
nk_BT/B^2$ is the ratio of the thermal and magnetic pressures).
Observations of reconnection at the terrestrial magnetopause
\citep{scurry94a} and in the solar wind \citep{phan09a} suggest that,
during high $\beta$ conditions, reconnection only occurs between
anti-parallel magnetic fields, i.e., when the shear angle between the
reconnecting fields is $\approx 180^{\circ}$.  Particle-in-cell (PIC)
simulations of reconnection have demonstrated why this occurs:
diamagnetic drifts in plasmas with a high $\beta$ but shear angles
$<180^{\circ}$ suppress reconnection \citep{swisdak03a}.

In this paper we present PIC simulations showing that diamagnetic
stabilization also occurs at the HP and limits the possible
sites of reconnection to those where the interstellar magnetic field
lies nearly anti-parallel to the heliospheric field.  Large-scale MHD
simulations of the heliosphere can then determine where HP
reconnection should occur.

We find the locus of anti-parallel reconnection sites to be relatively
small, so in situ detection of reconnection signatures at the HP by
either Voyager spacecraft seems unlikely.  However, on several
occasions since their launch, the spacecraft have detected bursts of
radio emission at 2-3 kHz believed to be associated with the
interaction of interplanetary shocks and the HP \citep{gurnett03a}.
We argue that reconnection triggered by this interaction drives the
electron beams producing the radio emission.  By combining the
locations determined for one set of bursts \citep{kurth03a} with the
requirement that HP reconnection be nearly anti-parallel we can
constrain the magnitude and orientation of the local interstellar
magnetic field (LISM).  In particular, the vector direction of the
LISM (the difference between $\mathbf{B}$ and $-\mathbf{B}$) can be
easily determined.  This is, to our knowledge, the first determination
of this parameter.

\section{Heliospheric Model}\label{MHD}

As part of our investigation we perform large-scale 3D MHD simulations
of the heliosphere.  The code is based on BATS-R-US, a
three-dimensional parallel, adaptive grid code developed by the
University of Michigan \citep{gombosi94a} and adapted for the outer
heliosphere \citep{opher04a}.

The Sun is the center of our coordinate system with the $x$ axis
parallel to the line through, but pointing inward from, the nose of
the HP.  The $z$ axis lies parallel to the solar rotation axis and the
$y$ axis completes the right-handed triplet.  In order to better
describe the different sub-environments of the heliosphere such
simulations model multiple fluids \citep{zank96b,pogorelov06a}; we
include five hydrogen populations, one ionized and four neutral
\citep{opher09a}.  All four neutral populations are
described by separate systems of Euler equations with corresponding
source terms describing neutral-ion charge exchange.  The parameters
for the inner boundary (located at 30 AU) were chosen to match those
used by \cite{izmodenov08a}: proton density at the inner boundary of
$n=8.74\times 10^{-3}\,\text{cm}^{-3}$, temperature $T=1.087\times
10^5\,\text{K}$, and a Parker spiral magnetic field with strength $B =
2\mu\text{G}$ at the equator.  Except where otherwise noted the
heliospheric field orientation corresponds to that during solar cycle
22 (roughly 1986 to 1996), in which in the $x-z$ plane through the nose
of the heliopause the field in the northern hemisphere points in the
$-\mathbf{\hat{y}}$ direction. We denote this as $B_{SW,y}<0$.

The outer boundary conditions are $n=0.06\,\text{cm}^{-3}$, velocity
$=26.3\,\text{km/s}$, and $T=6519\,\text{K}$.  The neutral hydrogen in
the LISM is assumed to have $n=0.18\,\text{cm}^{-3}$ and the same
velocity and temperature as the ionized LISM.  We use fixed inner
boundary conditions for the ionized fluid and soft boundaries for the
neutral fluids.  We impose outflow outer boundary conditions
everywhere except the $-\mathbf{\hat{x}}$ boundary where inflow
conditions are used for the ionized and neutral populations from the
interstellar medium.  The grid's outer boundaries are at $\pm 1500$ AU
in all three directions and the computational cell sizes range from
$0.73$ to $93.7$ AU.  The orientation of the interstellar magnetic
field is characterized by the angles $\alpha_{IS}$, the angle between
the field and the interstellar wind, and $\beta_{IS}$, the the angle
between the field and the solar equator.  Given $\alpha_{IS}$ and
$\beta_{IS}$, the vector direction of the field is still undetermined
up to a sign that can be fixed by noting whether the $y$ component of
$\mathbf{B}_{ISM}$ is positive or negative.

\begin{figure}
\plotone{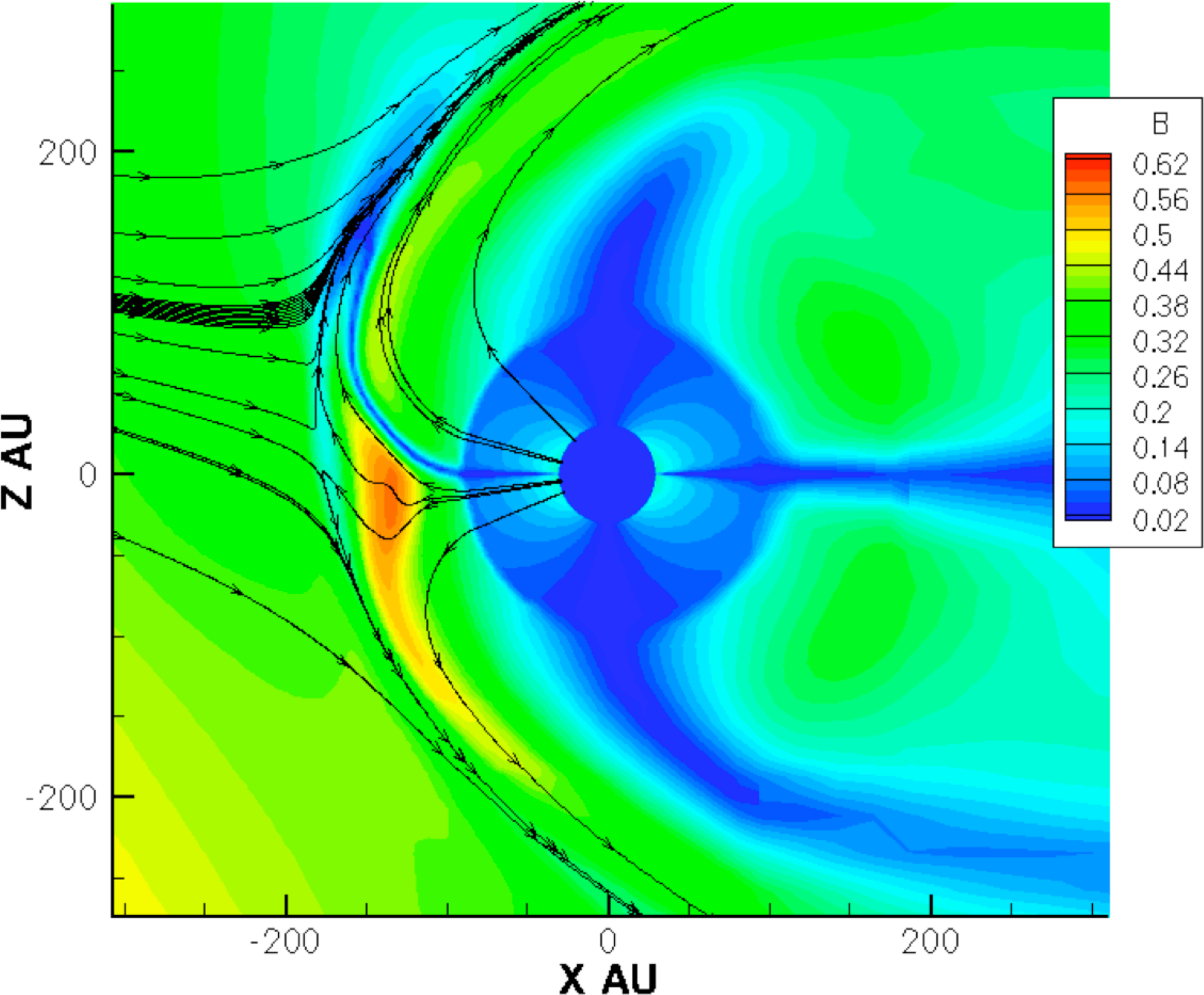}
\caption{\label{overview} Cut in the $x-z$ plane of a 3D MHD simulation
of the outer heliosphere with $B_{ISM,y}>0$ and a heliospheric field
with the solar cycle 22 polarity ($B_{SW,y}<0$).  Colors denote
magnetic field strength (in nT) and the black lines represent flow
streamlines.  The HP traces the outer edge of the the heliospheric
current sheet, which is shown by the blue line paralleling the
$-\mathbf{\hat{x}}$ axis and then deflecting northward in the
heliosheath.}
\end{figure}

In Figure \ref{overview} we show an overview of a heliospheric
simulation with $B_{ISM,y} > 0$ and the heliospheric polarity of solar
cycle 22.  The colors represent $|\mathbf{B}|$ and the black lines
are streamlines of the flow.  The roughly circular surface at a radius
of 80 AU from the Sun where the magnetic field strength increases from
$\approx 0.02$nT to $\approx 0.2$nT is the termination shock.
Downstream from the shock the plasma density and temperature increase
and the solar wind velocity decreases, as expected.  The neutral
hydrogen species do not have a dramatic effect on the overall
morphology of the system, although they affect the distances to the
termination shock and HP, bringing them closer to the Sun, as can be
seen by comparison to previous simulations employing only the ionized
fluid \citep{opher06a,opher07a}.

Reconnection occurs at the grid scale at a rate determined by
numerical details rather than by physical processes in these MHD
simulations.  Previous modeling has shown that without the addition of
some sort of extra-physical diffusive process, e.g., an anomalous
resistivity, MHD reconnection is inherently slow \citep{biskamp86a}.
The inclusion of the Hall term in the generalized Ohm's law
dramatically enhances the reconnection rate \citep{birn01a}, but
necessitates the resolving of length scales on the order of the ion
inertial length $d_i = c/\omega_{pi}$.  For a typical heliopause
density of $0.05 \text{ cm}^{-3}$, $d_i \approx 10^8 \text{ cm}
\approx 10^{-5} \text{ AU}$, several orders of magnitude smaller than
the resolution of the global simulation.  Since the MHD model does not
correctly describe reconnection, we use it as a starting point for
kinetic simulations of HP reconnection.

\section{Kinetic Reconnection Simulations}\label{diamag}

\subsection{Numerical Methods}\label{nummethods}
For our kinetic simulations we use p3d, a massively parallel PIC code
\citep{zeiler02a}.  We use a different coordinate system than the MHD
simulations because our computational domains represent only a tiny,
arbitrarily oriented, portion of the MHD domain.  The inflow into and
outflow from an X-line lie parallel to $\mathbf{\hat{y}}$ and
$\mathbf{\hat{x}}$, respectively, while the reconnection electron
field and any magnetic guide field parallel $\mathbf{\hat{z}}$.  In
the simulations presented here we assume out-of-plane derivatives
vanish, {\it i.e.}, $\partial/\partial z = 0$.  Although this choice
eliminates any structure in the $\mathbf{\hat{z}}$ direction, previous
studies indicate the basic features of reconnection remain unchanged
\citep{hesse01a}.

Masses are normalized to the ion mass $m_i$, magnetic fields to the
asymptotic value of the reversed field $B_0$, and the density to the
value at the center of the current sheet.  Other normalizations derive
from these: velocities to the Alfv\'en speed $v_A$, lengths to the ion
inertial length $d_i = c/\omega_{pi}$ (with $\omega_{pi}$ the ion
plasma frequency), times to the inverse ion cyclotron frequency
$\Omega_{ci}^{-1}$, and temperatures to $m_i v_A^2$.

We set the speed of light (in normalized units) to $20$ and the ratio
of the ion and electron masses to $m_i/m_e = 100$.  The spatial
resolution is such that there are $>4$ gridpoints per electron
inertial length and $\approx\negmedspace2$ per Debye length.  The
Courant condition determines the particle timestep and we substep the
advancement of the electromagnetic fields.  A typical cell contains
$\sim\negmedspace100$ particles and our simulations follow
$>\negmedspace10^9$ particles.  All of our runs conserve energy to
better than $1$ part in $100$.

\subsection{Simulation Results}

The components of any plasma that includes both a magnetic field and a
non-parallel pressure gradient undergo a diamagnetic drift given by
\begin{equation}\label{vstar}
\mathbf{v}_{*,j} = -c\frac{\bm{\nabla} p_j \bm{\times}
\mathbf{B}}{q_jn_jB^2}\text{,}
\end{equation}
where $p_j=n_jk_BT_j$ is the pressure and $q_j$ is the charge of
species $j$.  Note that, because of their charges, ions and electrons
drift in opposite directions and that separate populations of the same
species (in particular, pickup and thermal ions) can drift at
different speeds.  In our case the pressure gradient normal to the HP
crossed with the guide magnetic field produces drifts parallel to the
reconnection outflows.

Using PIC simulations \cite{swisdak03a} demonstrated that under such
conditions the X-line convects in the ion rest frame with a speed
given by $|\mathbf{v}_{*e}| + |\mathbf{v}_{*i}|$, the sum of the
electron and ion diamagnetic drifts.  If the drift velocity of the
X-line exceeds the speed of the nominally Alfv\'enic outflows,
reconnection is suppressed.  Qualitatively, suppression occurs when
the X-line propagates fast enough that the plasma does not have
sufficient time to establish the necessary flow configuration for
reconnection before the X-line passes.  Quantitatively,
\cite{swisdak03a} proposed that diamagnetic drifts suppress
reconnection when
\begin{equation}\label{condition1}
v_{*,j} > v_{A,j}
\end{equation}
where $v_{A,j}$ is the Alfv\'en speed for a particular populations
(i.e., calculated based on $n_j$) corresponding to the reconnecting
magnetic field.  This condition can be derived from balancing the
centrifugal and magnetic tension forces exerted on a fluid streaming
with velocity $v_{*,j}$ towards the X-line (see the Appendix for
details).  When $v_{*,j}$ is sufficiently large the magnetic tension
cannot eject the field line from the vicinity of the X-line and
reconnection stalls.

Equation \ref{condition1} can be reformulated as a condition relating
the jump in the plasma parameter across the current layer,
$\Delta\beta$, and the shear angle $\theta$ between the reconnecting
fields:
\begin{equation}\label{condition2}
\Delta\beta_j > \frac{2L_p}{d_i}\tan(\theta/2)
\end{equation}
where $L_p$ represents a typical pressure scale length near the
X-line.  Both simulations and observations at the Earth's magnetopause
suggest that $L_p/d_i \sim \mathcal{O}(1)$
\citep{berchem82a,eastman96a}.  According to equation
\ref{condition2}, only anti-parallel reconnection, $\theta \approx
\pi$, occurs for $\beta
\gg 1$.  Observations of reconnection at the magnetopause
\citep{scurry94a} and in the solar wind \citep{phan09a} support this
conclusion.  

\begin{figure}
\plotone{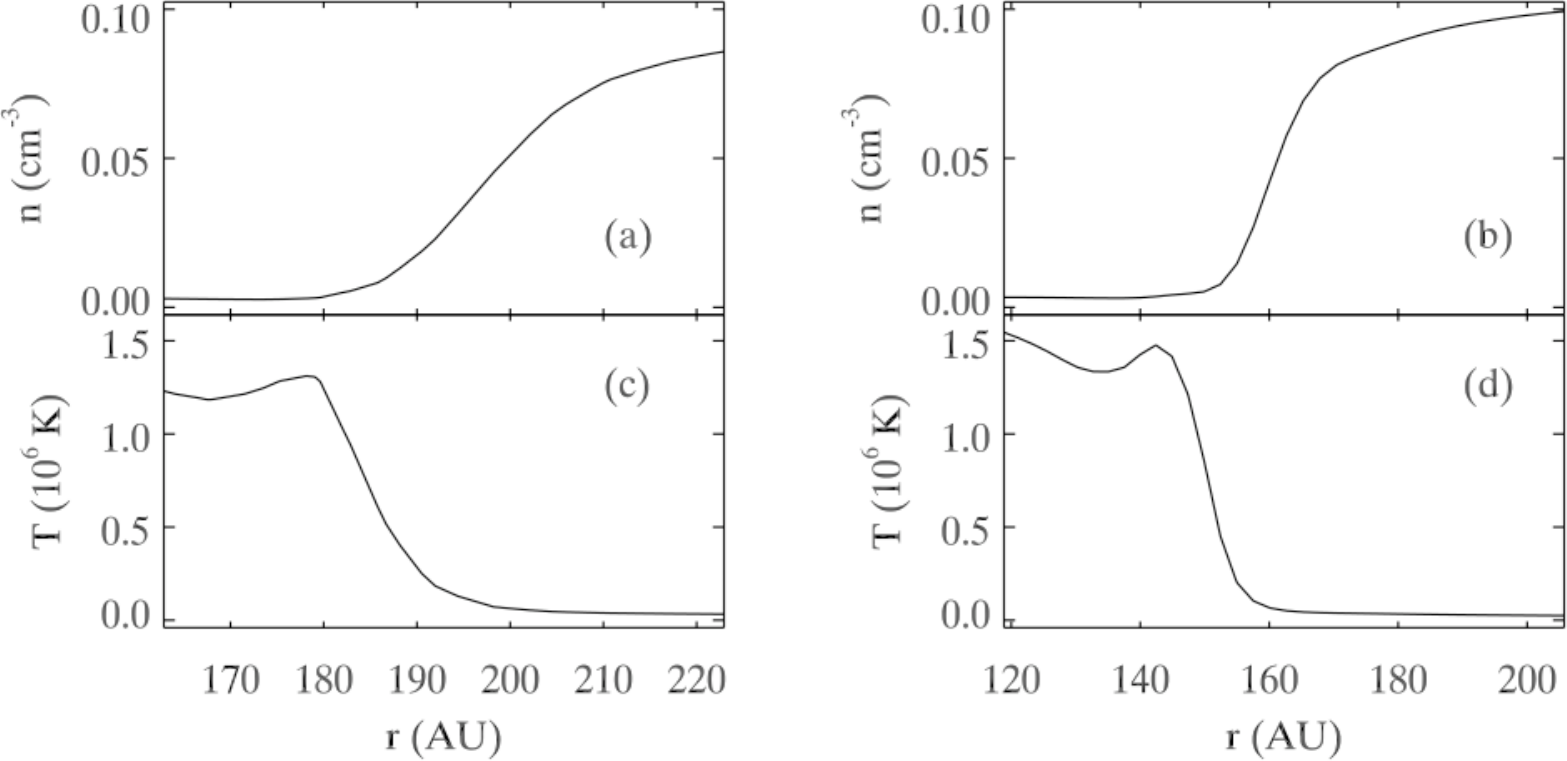}
\caption{\label{init} Cuts taken through the HP from the MHD
simulation showing the density and pickup ion temperature.  Panels (a)
and (c) come from a site, location 1, with anti-parallel reconnection;
panels (b) and (d) from location 2 where the fields have a smaller
shear angle.}
\end{figure}

The inferred large energy content of pickup ions in the heliosheath
implies a substantial value for $\beta$, which, when coupled with
equation \ref{condition2}, leads to the conclusion that only
anti-parallel reconnection occurs at the HP.  To check, we performed
PIC simulations of HP reconnection at two shear angles.  Cuts through
the HP at two places in the simulation of Figure
\ref{overview} established the initial equilibria for these
simulations.  Due to the relatively large grid size the heliopause in
the global simulations has a thickness of approximately 10 AU.  This
is only an upper limit, however, as the real thickness is certainly
closer to the $1 d_i \ll 10 \text{AU} $ observed at the terrestrial
magnetopause.  Furthermore, in both experiments
\citep{yamada07a} and simulations \citep{cassak05a} reconnection is
seen to proceed slowly in thick current sheets before accelerating
rapidly when the sheet thickness reaches $d_i$.  In our kinetic
simulations the MHD models provide the asymptotic parameters of the
reconnecting plasmas, but we take the width of the current layer to be
$1 d_i$. In Figure \ref{init} we show the initial profiles of the
densities and temperatures at the two locations used to provide the
asymptotic profiles for the PIC simulations.

The initial proton distribution is a superposition of a cold,
Maxwellian population representing the solar wind and a much hotter
Maxwellian (20\% by number) corresponding to the pickup ions. Choosing
a thermal distribution for the pickup ion distribution is a somewhat
crude approximation.  The actual distribution is still a subject of
investigation since neither of the Voyager spacecraft can measure it
directly.  Close to the Sun, where many of the pickup ions are
created, a ring distribution could be appropriate, but as the solar
wind expands outward a number of altering effects have been identified
\citep{isenberg87a,chalov95a} even before the interaction with the
termination shock introduces further modifications \citep{zank10a}.
In our kinetic simulations the particle distribution functions are
free to evolve (and do so, particularly in the reconnecting current
layer) and so the effects of choosing a different initial distribution
are ultimately unclear, albeit of obvious interest.

The first site has nearly anti-parallel reconnecting fields (shear
angle $\theta = 165^{\circ}$), while in location 2 the fields have a
shear angle of $\theta \approx 130^{\circ}$.  In order to specify the
plane of reconnection for these 2D simulations we use the prescription
derived in \cite{swisdak07a} for finding the X-line orientation based
on the asymptotic plasma parameters.  At locations 1 and 2 the values
of $\Delta\beta$ are $\approx 2$ and $4$, respectively.  Equation
\ref{condition2} then predicts that reconnection should be
diamagnetically stabilized at location 2, but not at location 1.

\begin{figure}
\plotone{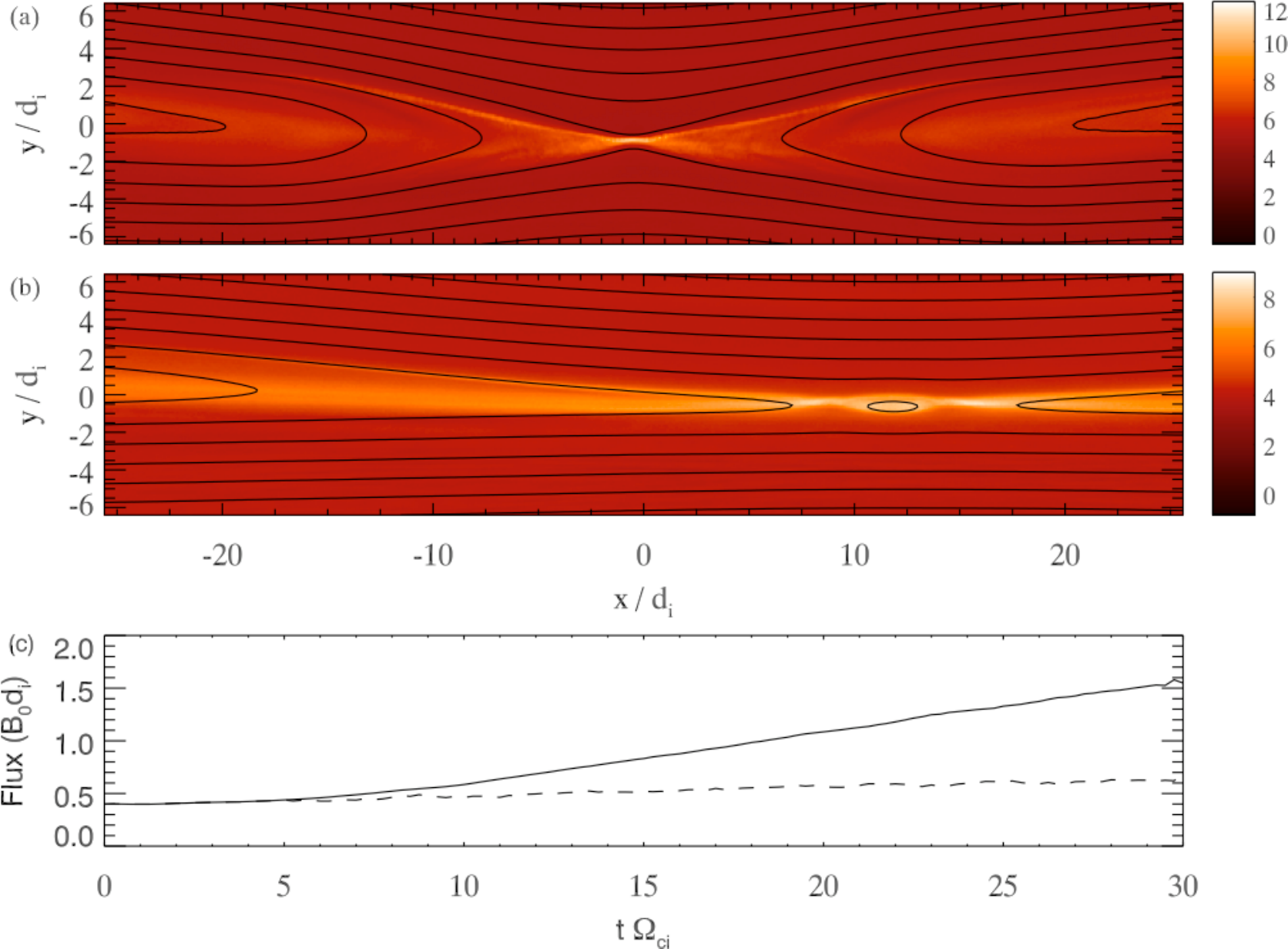}
\caption{\label{jz} Panels (a) and (b): Out-of-plane current density
overlaid by magnetic field lines for two PIC reconnection simulations.
The heliosheath and LISM plasma are above and below the current sheet
respectively.  Panel (a) corresponds to reconnection at location 1,
where the fields are anti-parallel.  Note the well-developed X-line.
In panel (b) we show reconnection at location 2, where the fields are
not anti-parallel.  The X-line, initially at $x/d_i = 0$, has drifted
due to diamagnetic effects.  Panel (c): Reconnected flux versus time
for the simulations shown in panels (a) (solid line) and (b) (dashed
line). The reconnection rate is given by the slope of each line.}
\end{figure}

In Figure \ref{jz}a we show a snapshot of the out-of-plane current
density from the anti-parallel case.  The bulge of the separatrices
into the low field strength heliosheath is typical of asymmetric
reconnection and has been observed in simulations of the terrestrial
magnetopause
\citep{kraussvarban99a,nakamura00a}.  Since the solar wind has the
lower field strength in both the terrestrial and heliospheric systems
the separatrices bulge toward the Sun in both cases.

We show a similar image from the non-anti-parallel reconnection case
in Figure \ref{jz}b.  Because of the diamagnetic drift the X-line has
drifted to the right from its initial position at approximately the
velocity given by equation \ref{vstar}.  A comparison of the size of
the islands downstream from the X-lines in the two simulations
indicates that essentially no flux has reconnected in Figure
\ref{jz}b.  The normalized reconnection rate as a
function of time for the two simulations can be seen in Figure
\ref{jz}c.  The anti-parallel case (solid line) clearly exhibits
reconnection at a rate of $\approx 0.06$, typical of ``fast'' Hall
reconnection \citep{birn01a}.  In the second case (dashed line), where
reconnection is not anti-parallel, very little flux reconnects.

\section{Radio Emission}

The plasma wave instruments on the Voyager spacecraft have detected
several bursts of radio emission at frequencies between 2 and 3 kHz
during their more than 25 years of operation.  A strong interplanetary
shock passed by the spacecraft before each burst, which led to the
suggestion that the interaction of global merged interaction regions
(GMIRs) and the HP boundary produced the emission \citep{gurnett06a}.
An application of several direction-finding techniques to one set of
bursts demonstrated that the sources lie at the radial distance of the
HP and stretch along a line parallel to the galactic plane and passing
near the nose of the heliosphere.

\cite{cairns02a} have suggested a detailed mechanism for producing the radio
emission.  Just beyond the HP, lower hybrid waves can resonantly
accelerate ambient electrons that then serve as a seed population for
shock acceleration.  The enhanced electron beams that form then excite
Langmuir oscillations, which subsequently decay, via nonlinear
effects, into the observed electromagnetic radiation.  This scenario
successfully explains, first, why the Voyager spacecraft do not
observe radio emissions (or the precursor Langmuir waves and electron
beams) when passed by shocks in the inner heliosphere, and second, why
the 2-3 kHz signal seems to be radially confined to a thin layer near
the HP.

Reconnection at the HP can also explain these observations.  Within
the heliosphere the current layers composing the heliospheric current
sheet are broad enough ($\sim 100 d_i$) that, despite occasional
observations of X-lines \citep{gosling07a}, reconnection proceeds
slowly on average.  Instead, the most significant reconnection should
occur when the plasma encounters a boundary, as happens at planetary
magnetopauses and, presumably, the HP.  The superposition of solar
wind structures in a GMIR means that the magnetic field strength can
be larger, by a factor of $\approx 4$, than the ambient value
\citep{whang95a}.  At the heliopause this enhanced field will produce
stronger reconnection and therefore naturally accounts for the
association between the arrival of the GMIRs at the heliopause and the
bursts of emission.  Second, observations and simulations have 
demonstrated that reconnection can produce both beams of electrons
traveling at the electron Alfv\'en speed \citep{cattell05a} and more
isotropic populations at relativistic energies
\citep{oieroset02a,hoshino01a,drake06a}.  The electromagnetic decay
mechanisms described in \cite{cairns02a} can then generate the 2-3 kHz
radiation from these electrons.  Reconnection, then, naturally
localizes the emission at the heliopause, produces the necessary
energetic electrons, and correlates the emission with the arrival of
GMIRs.

A significant test of this model is whether the locations of the 2-3
kHz radiation, as determined by \cite{kurth03a} for a series of events
between 1992 and 1994 (during solar cycle 22), correspond to the
regions where reconnection should occur.  The diamagnetic
stabilization of reconnection facilitated by pickup ions described in
Section \ref{diamag} implies that reconnection will only occur where
the reconnecting fields are anti-parallel.  The locations where such
reconnection occurs depend on the orientation of the interstellar
magnetic field, which will not be directly measured until one of the
Voyager spacecraft passes the heliopause.  However inferences from
indirect measurements, including backscattered Lyman-$\alpha$ emission
and the difference in the flow directions of interstellar hydrogen and
helium (the hydrogen deflection plane) do present some constraints
(see \cite{opher07a} and references therein).  Other constraints have
been derived based on the heliospheric asymmetries and heliosheath
flows measured by Voyager 2
\citep{opher09b}.  If reconnection is the source of the 2-3 kHz
radiation, the requirement that the sources lie near the locus of
anti-parallel reconnection can provide a strong further constraint on
the vector direction of the interstellar magnetic field.

We used the results of the global MHD model described in Section
\ref{MHD} to compare the locations of the radio sources determined by
\cite{kurth03a} and the locations of anti-parallel reconnection at the
HP.  We first define the HP in the simulation as the empirically
defined surface where $\log (T) = 10.9$; other values shift the HP
somewhat but do not significantly affect our conclusions.  Assuming
the field in the solar wind, $\mathbf{B}_{SW}$ is purely azimuthal
upstream of the HP, and hence neglecting any draping of the
heliospheric field, we calculate where
$\mathbf{B}_{SW}\boldsymbol{\cdot}(\mathbf{B}_{ISM}
\boldsymbol{\times}\mathbf{ \hat{n}})/|
\mathbf{B}_{ISM}||\mathbf{B}_{SW}|$ vanishes, where 
$\mathbf{B}_{ISM}$ is the field in the local interstellar medium and
$\mathbf{\hat{n}}$ is the normal to the HP.  (We use this metric
rather than the simpler
($\mathbf{B}_{ISM}\boldsymbol{\cdot}\mathbf{B}_{SW})/|
\mathbf{B}_{ISM}||\mathbf{B}_{SW}| = -1$ to eliminate the effects of
spurious normal magnetic fields at the heliopause.)  This quantity
vanishes for both anti-parallel and parallel field configurations so
we must combine it with another measure (we use the field shear angle)
to identify the anti-parallel locations.

\begin{figure}
\begin{center}
\epsscale{0.5}
\plotone{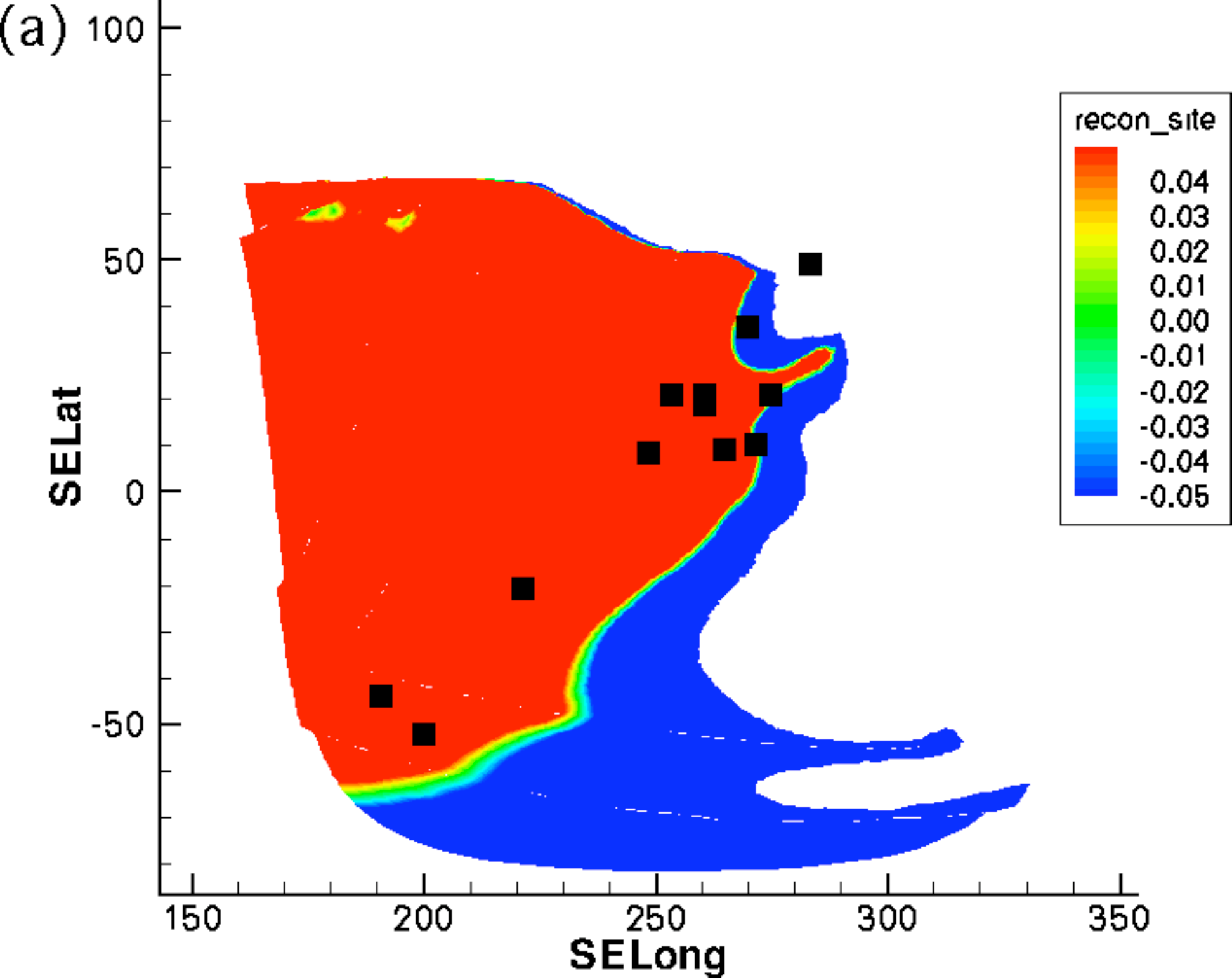}
\end{center}
\epsscale{1.}
\plottwo{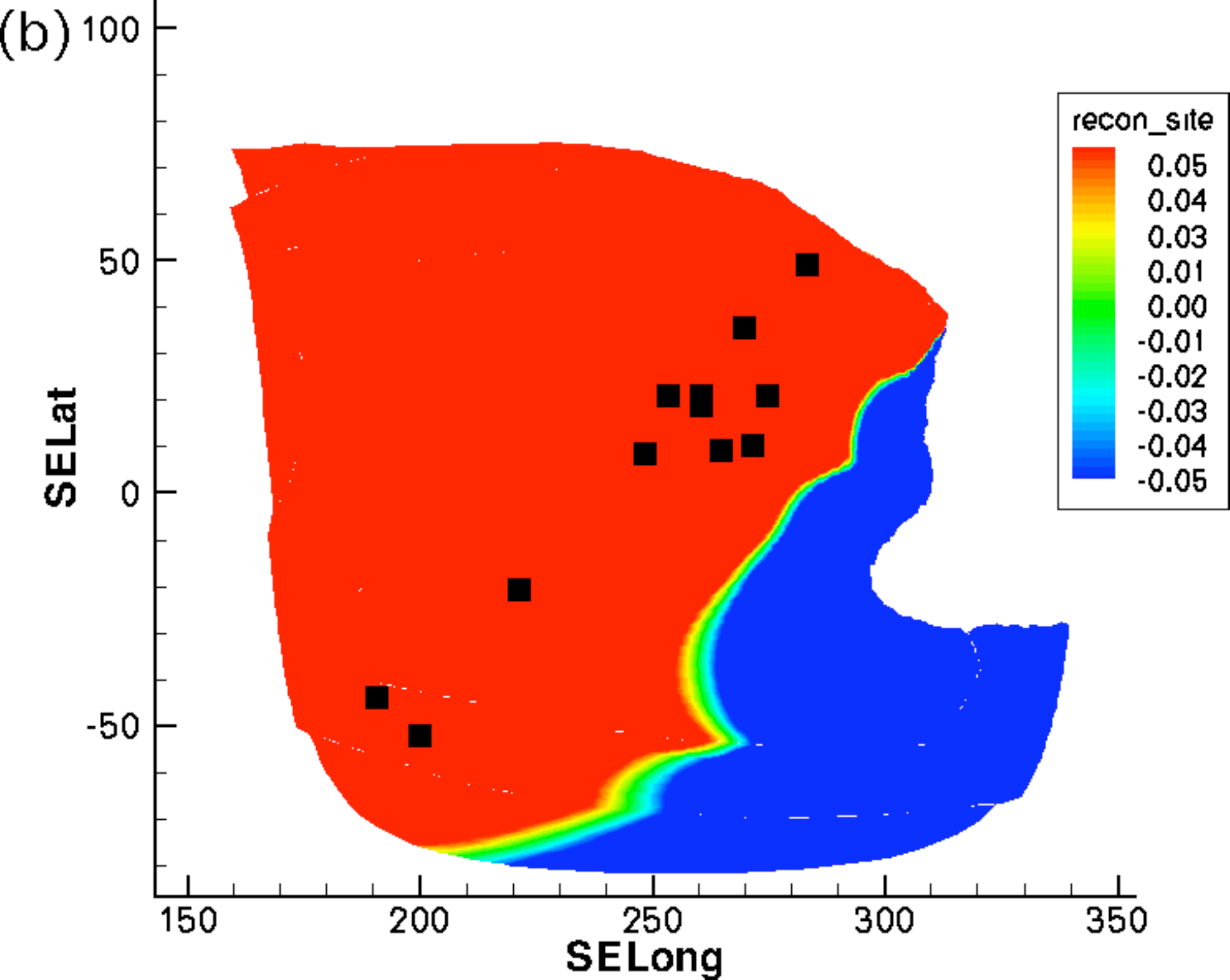}{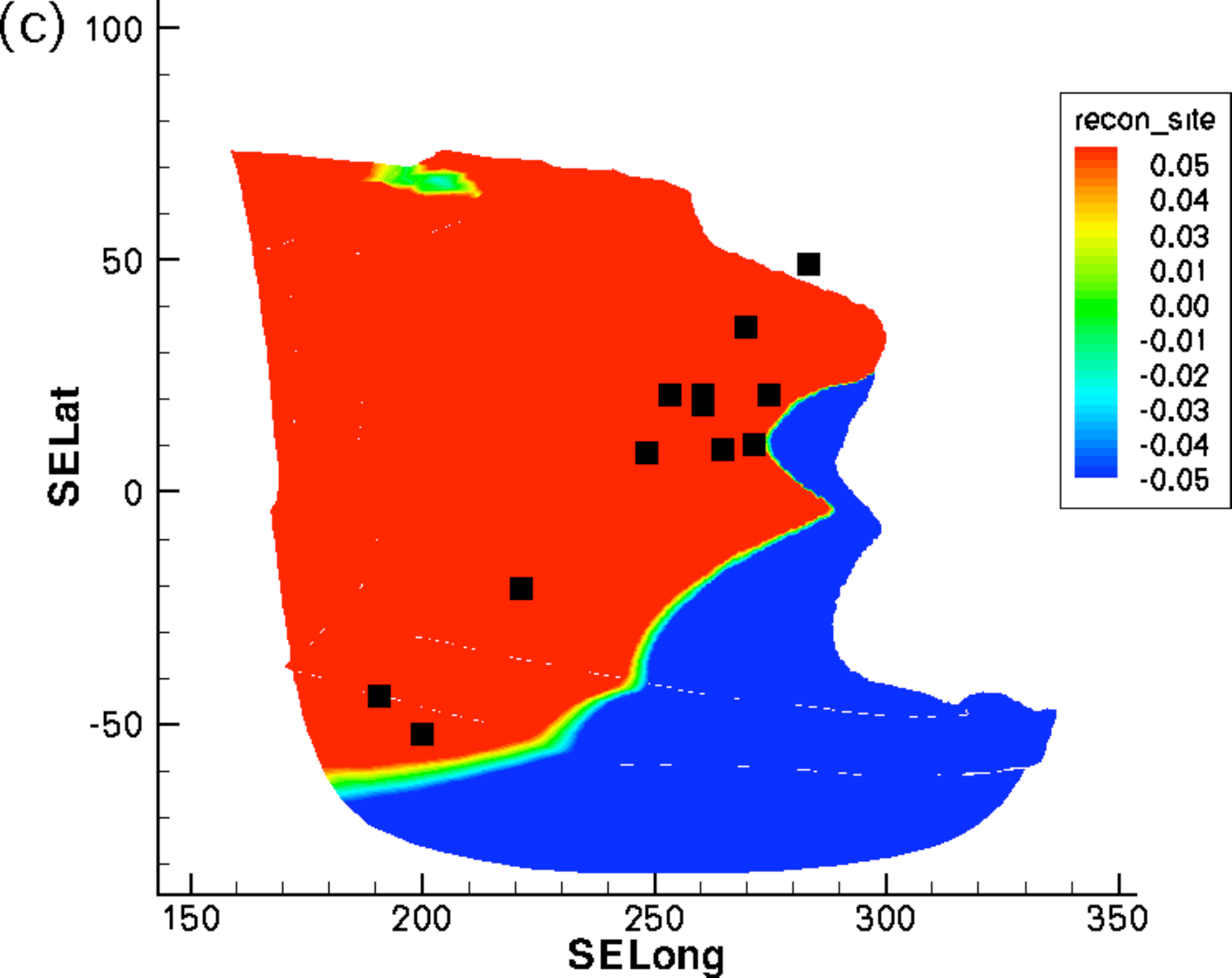}
\caption{\label{radiosources} Locations of anti-parallel reconnection
at the HP for three configurations of the interstellar magnetic field
as seen from outside the heliosphere looking inward.  SELat and SELong
are solar ecliptic latitude and longitude, respectively.  Colors
denote the quantity $\mathbf{B}_{SW}\boldsymbol
{\cdot}(\mathbf{B}_{ISM}
\boldsymbol{\times}\mathbf{ \hat{n}})/|
\mathbf{B}_{ISM}||\mathbf{B}_{SW}|$ and have been saturated at either
extreme.  White regions are places where the shear angle between
$\mathbf{B}_{SW}$ and $\mathbf{B}_{ISM}$ is less than $90^{\circ}$.
The squares represent the sources of radio emission, as determined by
\cite{kurth03a}. In panel (a) $\beta_{IS}=60^{\circ}$ and
$\alpha_{IS}=30^{\circ}$, in panel (b) $\beta_{IS}=80^{\circ}$ and
$\alpha_{IS}=30^{\circ}$, and in panel (c) $\beta_{IS}=60^{\circ}$ and
$\alpha_{IS}=20^{\circ}$.  In all panels $B_{ISM,y}>0$ and
$B_{SW,y}<0$ (solar cycle 22 polarity).}
\end{figure}

In Figure \ref{radiosources}a we show the map produced by the
interstellar field parameters --- $\alpha_{IS} = 30^{\circ}$,
$\beta_{IS} = 60^{\circ}$, $|\mathbf{B}| = 4.4\,\mu\text{G}$,
$B_{ISM,y}>0$ --- that provide the best fit.  These parameters are
consistent with the values determined by other methods.
%It is important to note that the locations of
%several of the radio sources have significant uncertainties.  In
%particular, \cite{kurth03a} suggest that scattering of the emission in
%the inner heliosphere can mask the true location and make bursts
%appear as omni-directional, rather than point, sources.  In Figure
%\ref{radiosources} the sources with the most uncertain locations are
%shown in gray.  
Our models suggest that the locus of anti-parallel
reconnection is more sensitive to $\alpha_{IS}$ and $|\mathbf{B}|$
than it is to $\beta_{IS}$.  Changing $\alpha_{IS}$ by $10^{\circ}$ or
$|\mathbf{B}|$ by 20\% moves the anti-parallel locus significantly
away from the sources, while shifts in $\beta_{IS}$ of up to
$20^{\circ}$ keep the concordance acceptable.  Panels
\ref{radiosources}b and \ref{radiosources}c displays the map for
$\alpha_{IS} = 30^{\circ}$, $\beta_{IS} = 80^{\circ}$ and $\alpha_{IS}
= 20^{\circ}$, $\beta_{IS} = 60^{\circ}$, parameters at the edge of
what we deem the acceptable range.

Perhaps of most interest is the dependence of our results on the
vector sign of the local interstellar magnetic field.  Figure
\ref{radiosources} corresponds to an interstellar field with $B_{ISM,y} > 0$.
Reversing the sign of $\mathbf{B}_{ISM}$ leads to a heliopause with no
anti-parallel reconnection sites near the nose.  If anti-parallel
reconnection\footnote{Although diamagnetic stabilization provides one
way, any mechanism that prevents non-anti-parallel reconnection will
have the same effect.} causes the 2-3 kHz radio bursts at the
heliopause, then the existence of these bursts fixes the vector
direction of the interstellar magnetic field.

The polarity of the heliospheric field reverses due to the 11-year
solar cycle.  A location that exhibits anti-parallel reconnection
during one cycle will, in the next 11 years, find that the
heliospheric and interstellar fields are parallel, and hence unlikely
to reconnect\footnote{Note that in the MHD code reversing the
orientations of both the interstellar and heliospheric fields has no
effect on the loci of anti-parallel reconnection: $B_{y,ISM} > 0$ with
a given orientation of the heliospheric field behaves identically to
$B_{y,ISM} < 0$ with the opposite heliospheric orientation.}.  A
similar change happens at the Earth's magnetosphere when the
interplanetary magnetic field rotates (on a much faster time scale)
from, for instance, southward to northward.  However, just as in the
magnetospheric case, even when anti-parallel reconnection is not
possible near the nose of the HP it can occur, albeit more weakly, at
much higher latitudes near the cusps of the Earth's field.  This
effect cannot be directly tested with the data from Figure
\ref{radiosources} because all of the source locations were determined
during a two-year period (1992--1994) of solar cycle 22. The Voyager
spacecraft did detect radio bursts during solar cycles 21 and 23, but
could not determine the locations of their sources, specifically
if they occurred at the cusps as our model suggests.  Intriguingly,
these events were substantially weaker than those of solar cycle 22,
which agrees with the observation that in the terrestrial case cusp
reconnection is weaker than its equatorial counterpart.

\begin{figure}
\plotone{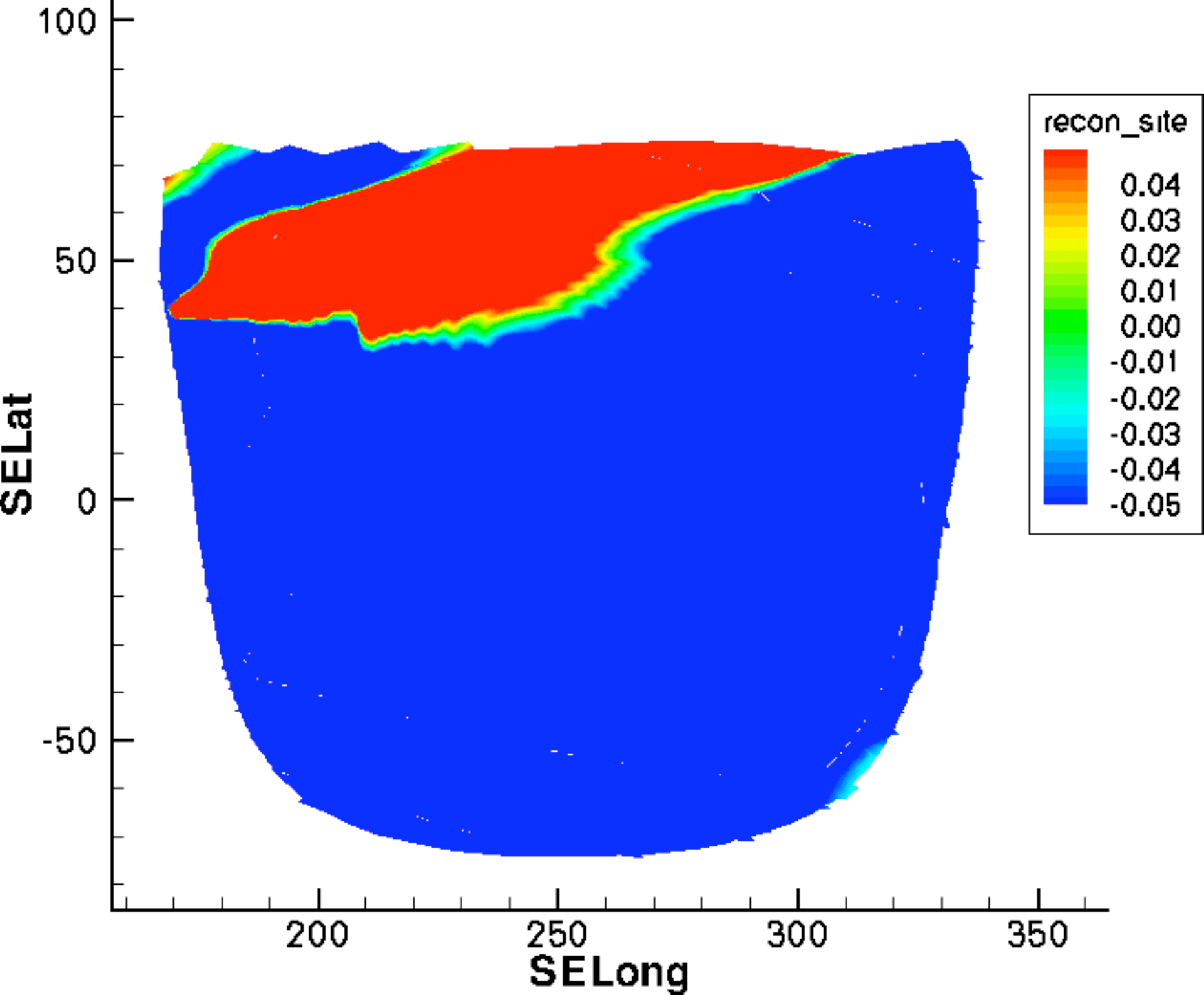}
\caption{\label{currentdipole} Angles between the interstellar and
heliospheric fields for the solar cycle 23 orientation of the solar
dipole.  As in Figure \ref{radiosources} colors denote
$\mathbf{B}_{SW}\boldsymbol {\cdot}(\mathbf{B}_{ISM}
\boldsymbol{\times}\mathbf{ \hat{n}})/|
\mathbf{B}_{ISM}||\mathbf{B}_{SW}|$, but here the fields have shear
angles $< 90 ^{\circ}$.  For this run, $\beta_{IS}=60^{\circ}$,
$\alpha_{IS}=20^{\circ}$.}
\end{figure}

Figure \ref{currentdipole} displays results for the solar cycle 23
polarity of the solar dipole ($B_{SW,y} >0$) for a case otherwise
identical to that shown in Figure \ref{radiosources}c ---
$\beta_{IS}=60^{\circ}$, $\alpha_{IS}=20^{\circ}$, $|\mathbf{B}| =
4.4\,\mu\text{G}$, $B_{ISM,y}>0$.  If plotted in the same manner as
Figure \ref{radiosources} the entire figure would be white because the
interstellar and heliospheric fields have shear angles $<90^{\circ}$.
Instead the figure shows that $\mathbf{B}_{SW}$ and $\mathbf{B}_{ISM}$
are nearly parallel over the entire face of the heliopause.  In this
case we expect anti-parallel reconnection to only happen at high
heliospheric latitudes, however the inadequate resolution of the
global models there does not allow us to accurately predict the
favorable locations.

Future data sources may allow us to test this dependence.  This model
predicts a transition to energetic reconnection near the nose in solar
cycle 24 as the polarity of the heliospheric current sheet reverses
again.  However current observations suggest that the Sun may be
entering a deep solar minimum, with very little magnetic activity.  If
true it may be accompanied by a concomitant decrease in heliopause
reconnection.  In any case, in situ samplings of reconnection by the
Voyager spacecraft, aside from being unlikely due to the small odds
that a given outward trajectory through the heliopause will pass near
an X-line, will not be possible for at least another decade.  Indirect
measurements of reconnection at the HP are more likely in the near
future.  Observations \citep{lin03a} and simulations
\citep{drake09b} have demonstrated that reconnection can energize ions
(this occurs in addition to the electric energization necessary for
the 2-3 kHz radio emission).  When these energetic ions interact with
the surrounding plasma they can undergo charge exchange and create
energetic neutrals that can perhaps be sensed remotely by such
missions as IBEX.  Such a signal would appear in addition to the band
of energetic neutrals recently reported by IBEX \citep{mccomas09a}
which is, in this view, unrelated to HP reconnection.

%Finally, the heliospheric magnetic field includes a well-known sector
%structure arising from the misalignment of the solar rotation and
%magnetic axes \citep{ness65a}.  As the solar wind plasma approaches
%the HP, however, the embedded current sheets narrow, leading to a
%burst of reconnection and the formation of multiple magnetic islands
%\citep{drake09a}.  Although the narrowing at the heliopause has not
%been directly observed, an analogous effect occurs when solar wind
%current sheets encounter the terrestrial bow shock and magnetopause
%\citep{phan07a}.  The extent of the sectors in latitude varies
%considerably, reaching its peak during solar maximum.  Multiple
%reconnection events within the current sheet will tend to dissipate
%the sheet and destroy the overall structure of the magnetic field. So,
%even without diamagnetic stabilization we might expect reconnection
%with the interstellar magnetic field at the HP to be less important
%near the equator.  At higher latitudes, however, ($\gtrsim \pm
%30^{\circ}$), where the sector structure does not matter, the field
%will retain its continuity.  The regions of anti-parallel reconnection
%identified in Figure \ref{radiosources} lie on the outskirts, and
%possibly completely outside, of the sector structure.

\acknowledgments{We would like to thank the reviewer for helpful
comments that led to significant improvements in the paper.
Computations were carried out at the NASA AMES Research Center and
National Energy Research Supercomputing Center.  M. O. and
F. A. B. acknowledge the support of NASA-Voyager Guest Investigator
grant NNX07AH20G and the National Science Foundation CAREER grant
ATM-0747654.}

\appendix

\section{Condition for Diamagnetic Suppression of Reconnection}

To more formally derive the condition for diamagnetic suppression of
reconnection given in equation \ref{condition1}, begin by considering
the outflow from an X-line.  In the simplest case, with no diamagnetic
drifts, reconnection produces bent magnetic field lines that
accelerate away from the X-line due to the
$\mathbf{J}\boldsymbol{\times}\mathbf{B}$ force.  Specifically, in the
coordinate system described in Section \ref{nummethods} where outflow
is parallel to $\pm\mathbf{\hat{x}}$, the force is proportional to
$J_zB_y$.

In the case with diamagnetic drifts along the $\mathbf{\hat{x}}$ axis
one of the outflows from the X-line will be in the direction opposite
to the drifting plasma.  The change in momentum that a bent field line
can cause in a time $\Delta t$ over a box of dimensions $\Delta_x$ by
$\Delta_y$ is
\begin{equation}
J_zB_y\Delta_x\Delta_y\Delta t/c
\end{equation}
The unbending of the field line, and hence reconnection, can only
occur if this quantity is large enough to overcome the momentum of the
plasma in the box traveling with the diamagnetic velocity
\begin{equation}
-\rho_j v_{*,j}\Delta_y (v_{*,j}\Delta t)
\end{equation}
where $\rho_j$ and $v_{*,j}$ are the density and diamagnetic velocity
of species j.  Combining these expression leads to a condition for the
suppression of reconnection.  By using Amp\'ere's law to substitute for
$J_z = (\partial B_x/\partial y ) /(4\pi/c)$ and Gauss's law to equate
$\partial B _x/\partial x$ and $-\partial B_y/\partial y$ we arrive at the
suppression condition:
\begin{equation}\label{finalappeq}
\rho_j v_{*,j}^2 > \frac{B_x^2}{4\pi}
\end{equation}
Equation \ref{finalappeq} is equivalent to equation \ref{condition1}
of the main text.

%\begin{figure}
%\plotone{rate.pdf}
%\caption{\label{rate} Reconnected flux versus time for two PIC simulations.
%Anti-parallel reconnection at location 1 (solid line) exhibits steady
%reconnection while sheared reconnection at location 2 (dashed line) is
%stabilized due to the diamagnetic drift of the X-line.  The
%reconnection rate is give by the slope of each line.}
%\end{figure}

\bibliographystyle{agu04}
\bibliography{paper}

\end{document}